

 at 17.3 truept    \font\bgg=cmbx10 at 12 truept
\font\twelverm=cmr10 scaled 1200    \font\twelvei=cmmi10 scaled 1200
\font\twelvesy=cmsy10 scaled 1200   \font\twelveex=cmex10 scaled 1200
\font\twelvebf=cmbx10 scaled 1200   \font\twelvesl=cmsl10 scaled 1200
\font\twelvett=cmtt10 scaled 1200   \font\twelveit=cmti10 scaled 1200
\def\twelvepoint{\normalbaselineskip=12.4pt
  \abovedisplayskip 12.4pt plus 3pt minus 9pt
  \belowdisplayskip 12.4pt plus 3pt minus 9pt
  \abovedisplayshortskip 0pt plus 3pt
  \belowdisplayshortskip 7.2pt plus 3pt minus 4pt
  \smallskipamount=3.6pt plus1.2pt minus1.2pt
  \medskipamount=7.2pt plus2.4pt minus2.4pt
  \bigskipamount=14.4pt plus4.8pt minus4.8pt
  \def\rm{\fam0\twelverm}          \def\it{\fam\itfam\twelveit}
  \def\sl{\fam\slfam\twelvesl}     \def\bf{\fam\bffam\twelvebf}
  \def\mit{\fam 1}                 \def\cal{\fam 2}
  \def\tt{\twelvett}
  \textfont0=\twelverm   \scriptfont0=\tenrm   \scriptscriptfont0=\sevenrm
  \textfont1=\twelvei    \scriptfont1=\teni    \scriptscriptfont1=\seveni
  \textfont2=\twelvesy   \scriptfont2=\tensy   \scriptscriptfont2=\sevensy
  \textfont3=\twelveex   \scriptfont3=\twelveex  \scriptscriptfont3=\twelveex
  \textfont\itfam=\twelveit
  \textfont\slfam=\twelvesl
  \textfont\bffam=\twelvebf \scriptfont\bffam=\tenbf
  \scriptscriptfont\bffam=\sevenbf
  \normalbaselines\rm}
\def\doublespace{\baselineskip=\normalbaselineskip \multiply\baselineskip by 2}
\def\beginparmode{\endmode
  \begingroup \def\endmode{\par\endgroup}}
\let\endmode=\par
\newcount\equationnumber
\advance\equationnumber by1
\def\ifundefined#1{\expandafter\ifx\csname#1\endcsname\relax}
\def\docref#1{\ifundefined{#1} {\bf ?.?}\message{#1 not yet defined,}
\else \csname#1\endcsname \fi}
\def\autoeqnum{\def\eqlabel##1{\edef##1{\the\equationnumber}}}
\def\no{\eqno(\the\equationnumber){\global\advance\equationnumber by1}}
\newcount\citationnumber
\advance\citationnumber by1
\def\ifundefined#1{\expandafter\ifx\csname#1\endcsname\relax}
\def\cite#1{\ifundefined{#1} {\bf ?.?}\message{#1 not yet defined,}
\else \csname#1\endcsname \fi}
\def\autocite{\def\citelabel##1{\edef##1{\the\citationnumber}\global\advance\citationnumber by1}}
\def\preprintno#1{
 \rightline{\rm #1}}

\def\ss{\scriptscriptstyle}
\def\ef{_{\ss eff}}
\def\s{\scriptstyle}
\def\disp{\displaystyle}
\def\frac#1#2{{{#1}\over{#2}}}

\def\i{\infty}
\def\o{\over}
\def\l{\lambda}
\def\b{\beta}

\def\vf{\varphi}
\def\G{\Gamma}
\def\f{\phi}
\def\ft{\phi^2}
\def\k{\kappa}
\def\p{\partial}
\def\fb{\bar\f_B}
\def\gf{\gamma_{\f}}
\def\Zf{Z_{\phi}}
\def\gft{\gamma_{\f^2}}
\def\Zft{Z_{\phi^2}}
\def\Zl{Z_{\l}}
\def\gl{\gamma_{\l}}
\def\e{\varepsilon}
\def\de{d_{\ss eff}}

\hsize=6.5truein
\hoffset=.1truein
\vsize=8.9truein
\voffset=.05truein
\parskip=\medskipamount
\twelvepoint
\doublespace
\autocite
\autoeqnum

\vskip -48 truept
\preprintno{THU-93/18, DIAS-STP-93-18; August '93}
\vskip 12 truept
\centerline{\bgg EFFECTIVE CRITICAL EXPONENTS FOR}
\centerline{\bgg DIMENSIONAL CROSSOVER AND QUANTUM SYSTEMS FROM}
\centerline{\bgg AN ENVIRONMENTALLY FRIENDLY RENORMALIZATION GROUP}
\vskip\baselineskip
\centerline{\bf Denjoe O' Connor}
\centerline{School of Theoretical Physics,}
\centerline{Dublin Institute for Advanced Studies,}
\centerline{10 Burlington Road,}
\centerline{Dublin 4, Ireland.}
\vskip\baselineskip
\centerline{\bf C.R. Stephens }
\centerline{Institute for Theoretical Physics,}
\centerline{Rijksuniversiteit Utrecht, }
\centerline{Princetonplein 5, }
\centerline{3508TA Utrecht, Netherlands.}
\vskip 0.3truein
{\bf Abstract:}
Series for the Wilson functions of an ``environmentally friendly''
renormalization group are  computed to two loops, for an $O(N)$ vector model,
in terms of  the ``floating coupling'',
and resummed by the Pad\'e method to yield crossover exponents
for finite size and quantum systems. The
resulting effective exponents obey all scaling laws, including hyperscaling
in terms of an effective dimensionality, ${d\ef}=4-\gl$, which represents
the crossover in the leading irrelevant operator, and are in excellent
agreement with known results.
\vfill\eject
\beginparmode
Physical systems can exhibit different types of scaling behaviour in
different asymptotic regimes.  The crossover between such asymptotic
regimes is important both theoretically, and experimentally. One may
think of a crossover as being induced by some ``environmental''
variable.  Two of the most interesting crossovers are: those induced
by finite size effects
\citelabel{\Barbr}
[\cite{Barbr}], and those induced by quantum effects
\citelabel{\Hertz}[\cite{Hertz}], \citelabel{\ChkHaN}[\cite{ChkHaN}].
For finite size systems the environmental variable is $L$,
the system size, while in quantum systems it is the inverse absolute
temperature,  $\beta\hbar$. Other environmental factors that may
induce a crossover are: long range interactions, anisotropic
interactions, external fields, boundary conditions, etc..

The main difficulty in treating systems which exhibit a crossover is
that the qualitative nature of the effective degrees of freedom
(DOF), i.e. the fluctuations, changes significantly as a function of
scale, being very sensitive to the environment in the crossover
region. The renormalization group (RG) is our most powerful tool for
investigating how physical systems change as a function of scale.
If one views the RG transformation as a ``coarse graining''
procedure, one must ask whether a particular coarse graining is
capturing the qualitative changes associated with the crossover.
We call an RG which
tracks the changing nature of the effective DOF ---
``environmentally friendly''. Momentum shell integration is, in
principle, a naturally environmentally friendly form of
renormalization, however, the associated RG's are not easily
computed beyond
lowest order in perturbation theory. In addition one must be careful
not to throw away possible important environment dependence by
arguing that integrating out large momenta $k\gg g$, where $g$ is
the characteristic scale set by the environment, should be
$g$ independent, as this leads to RG flow equations which when
propagated to scales $k<g$ coarse grain effective DOF which are a
very poor representation of the system's fluctuations  at
that scale.

The most accurate RG results for properties of critical systems,
such as critical exponents, have been achieved by applying field
theoretic techniques
\citelabel{\BrLeGZnJ}
[\cite{BrLeGZnJ}], and are in very impressive agreement with
experiment
\citelabel{\BkNiM}
[\cite{BkNiM}]. Three approaches have been used:
$\varepsilon$ expansion
\citelabel{\WF}
[\cite{WF}], $1\o N$ expansion
\citelabel{\Abe}
[\cite{Abe}], and
fixed dimension perturbation theory
\citelabel{\Parisi}
[\cite{Parisi}]. The first fails in
crossovers where the upper critical dimension changes, such as in
finite size crossover. The second fails in crossovers where the
order parameter can change its symmetry, such as bicritical
crossover. We adopt the spirit of the
fixed dimension approach, if not the letter, in the context of
environmentally friendly renormalization.

Field theory historically, as distinct from momentum
shell integration, has emphasized the role of ultraviolet (UV)
divergences, which are independent of infrared scales and therefore
environment insensitive. The result is an RG which does not
track the changing nature of the effective DOF, and this leads,
typically, to a breakdown of perturbation theory.
More environmentally friendly RGs have been implemented in some
contexts. Amit and Goldscmidt
\citelabel{\AG}
[\cite{AG}] introduced the concept of
generalized minimal subtraction (GMS) when  considering crossover at
a bicritical point. Their results for $\gamma\ef$, however, differ
from those found using momentum shell integration
\citelabel{\NeDo}
[\cite{NeDo}],
the latter find a characteristic ``dip'' in the effective
exponent curves. Our methodology applied to a bicritical crossover
\citelabel{\DjCr}
[\cite{DjCr}] gives results in agreement with the momentum shell approach. GMS
was also applied to uniaxial dipolar ferromagnets in
\citelabel{\FSch}
[\cite{FSch}], the
results of our analysis
\citelabel{\Chr}
[\cite{Chr}, \cite{DjCr}] differ somewhat. Differences can be
understood, since our
renormalization procedure puts complete Feynman diagrams into the
Wilson functions, GMS does not. When the RG equation is solved all
the diagrammatic information in these functions is ``exponentiated'',
the remainder must be taken
into account perturbatively, in some way.
Other related work is that of Schmeltzer
\citelabel{\Schm}
[\cite{Schm}] who calculated
$\gamma\ef$ to one loop for a three dimensional quantum
ferroelectric and  Lawrie
\citelabel{\Lawrie}
[\cite{Lawrie}] who considered dimensional crossover
for $d$-dimensional quantal and $d+1$ dimensional finite-sized Ising
models for $3<d<4$ using an $\varepsilon$ expansion. Unlike our
method the $\varepsilon$ expansion  could not capture the
crossover between two non-trivial fixed points as
the upper critical dimension changes
across the crossover. Field theoretic results for
dimensional crossover in a fully finite geometry or a cylinder have
been obtained
\citelabel{\BrZnJ}
[\cite{BrZnJ}] but the techniques used have not been extended to
the case of a system with more than one fixed point.

Though our general approach is applicable to a very wide class of
crossovers
\citelabel{\EnvfRG}
[\cite{DjCr},\cite{EnvfRG}], we restrict our attention to finite size
crossover and quantum/classical crossover. We begin with the
``microscopic'' Landau-Ginzburg-Wilson Hamiltonian
\eqlabel{\ham}
$$H[\vf_{B}]=\int_0^L\int
d^{d}x\left[{1\o2}(\nabla\vf_{B})^2+{1\o2}m^2_{B}\vf_{B}^2+
{1\o2}t_{B}(x)\vf_{B}^2
+{\l_{B}\o4!}\vf_{B}^{4}-H_B(x)\vf_B\right]\no$$
which describes either: a layered $d+1$ dimensional system, of thickness $L$;
or a $d$ dimensional quantum system, with
$L=\beta\hbar$, $\beta$ being the inverse temperature. We
will assume the order parameter possesses an $O(N)$ symmetry, the
case $N=1$ of quantum/classical crossover represents an Ising
model in a transverse magnetic field. In the
finite size case $m_B^2+t_B=T-T_0$, and in the Ising model in a transverse
field $m_B^2+t_B=\G-\G_0$. Here, $T_0$ and $\G_0$ are the
critical temperature and transverse field respectively, in the mean field
approximation.

An $L$ dependent renormalization is necessary to obtain the desired
environmentally friendly RG. Use of $L$ dependent normalization
conditions achieves this and ensures that all the Feynman diagrammatic
information is exponentiated in the solution of the resulting RG equation.
The relation between the bare and renormalized vertex functions is
$\G_{B}^{(N,M)}=Z_{\f}^{-{N\o2}}Z_{\ft}^{-M}\G^{(N,M)}$
The renormalized dimensionful coupling is similarly related to the
bare one by
$\l_{B}=Z_{\l}^{-1}\l$  (see [\cite{BrLeGZnJ}] for a discussion of the
notation).
We choose the normalization conditions
\eqlabel{\normcnds}
$$\matrix{{\rm (i)}\quad \Gamma^{(2)}(k=0,t=\k^2,\l,L,\k)=\k^2&\qquad &
{\rm (ii)}\quad{\p\Gamma^{(2)}\over\p k^2}(k,t=\k^2,\l,L,\k)|_{\ss
k=0}=1\cr
& & \cr
{\rm (iii)}\quad\Gamma^{(4)}(k=0,t=\k^2,\l,L,\k)=\l&\qquad &
{\rm (iv)\;}\quad\Gamma^{(2,1)}(k,t=\k^2,\l,L,\k)=1\quad \cr}\no$$
which specify $Z_{\f}$,  $Z_{\ft}$   and $Z_{\l}$.
Condition (i), together with the multiplicative renormalization of
$t$, implies that $t$ is proportional to $T-T_c(L)$
for the finite size system, and
$\G-\G_c(\beta)$ for the quantal Ising model, i.e. that one is
measuring temperature/field deviations relative to the $L$ dependent
critical point. We are assuming here that the system can exhibit
critical behaviour for any value of $L$, which restricts our
attention to $d>1$ in the case of $N>1$, but in no way restricts
the generality of our approach however. We have applied our methods
successfully at one loop to dimensional crossover in a non-linear
$\sigma$ model
\citelabel{\Sigmod}
[\cite{Sigmod}] and find results in qualitative agreement with
those of [\cite{ChkHaN}] where a momentum shell integration approach was used,
also at the one-loop level. We believe our methods are more easily
extended to higher orders. If a normalization condition with $L=\i$
had been used, then temperature/field deviations would be measured
relative to $T_c(\i)$ or $\G_c(\i)$, the critical temperature of the
bulk system, or critical field of the $\beta=\i$ quantum system
respectively. Note that we are here using an RG which runs the renormalized
temperature parameter in distinction to the Callan-Symanzik equation which
runs the physical correlation length.

The RG equation can be viewed as a simple consequence of the fact that the
bare theory is independent of the arbitrary renormalization scale $\k$
at which we choose to define our parameters, i.e.
$\k{d\o d\k}\G^{(N)}_{B}=0$.
Using the relation between the bare
and renormalized vertex functions and expressing things in terms of
the renormalized parameters the infinitisimal form of the RG equation
then becomes
\eqlabel{\rge}
$$\left(\k{\p\o{\p\k}}+ \beta{\partial\over{\partial
\l}}+ \gamma_{\f^2}t{\partial\over {\partial t}} -{1\over2}{\gamma^{
}_{\f}}{\Bigl[} N+\fb{\partial\over{\partial \fb}}{\Bigl]}\right)
\Gamma^{\scriptscriptstyle{(N)}}=0\no$$
with
\eqlabel{\gfgftfl}
$$\gf={1\o\Zf}\k {d\Zf\o d\k};
\qquad\gft={1\o\Zft^{-1}}\k {d\Zft^{-1}\o d\k};
\quad{\rm and}\quad{\b(\l)\o\l}=\gl={1\o\Zl}\k{d\Zl\o
d\k}
\no$$
The functions $\gf$, $\gft$ and $\gl$ are the Wilson functions.
They are explicitly $L$ dependent due to the normalization
conditions (\docref{normcnds}) and all the physics of the
crossover can be gleaned from them.

A suitable coupling, with respect to which perturbation theory can be
performed, is the floating coupling
\citelabel{\NPhyOrig}
[\cite{NPhyOrig},\cite{DjCr}], $h$, which is chosen so as to make the quadratic
term in $\beta(h)$
have unit coefficient. Our perturbation theory is then carried out
at the level of the Wilson functions in terms of  $h$.
The expressions obtained are, however,
only the leading terms in an asymptotic expansion of the functions
$\beta(h,z)$, $\gft(h,z)$ and $\gf(h,z)$.
We use [2,1] Pad\'e approximants to resum these  asymptotic series obtaining
\eqlabel{\padebeta}
$$\beta(h,z)=-\e(z) h+{h^2\o 1+
4\left({(5N+22)\o {(N+8)}^{2}}f_{1}(z)
-{(N+2)\o{(N+8)}^{2}}f_{2}(z)\right) h}\no$$
and
\eqlabel{\padegft}
$$\gft(h,z)={(N+2)\o (N+8)}{h\o1+6{1\o{(N+8)}}\bigl(f_{1}(z)
-{1\o 3}f_{2}(z)\bigr) h}\no$$
where the functions $\e$, $f_1$ and $f_2$ depend on $d$ and
$z={\k L}$ but are independent of $N$. The original non Pad\'e resummed
series can be recovered by expanding $1/(1+x h)\sim 1-x h$.
We will take the solution of (\docref{padebeta}) as our perturbation
parameter. After these equations are solved it is then
inappropriate to do any further expansion.

The functions $\e(z)$, $f_1(z)$ and $f_2(z)$ are the basic building blocks,
their specific functional form depending on the particular crossover in
question. $\e(z)$
can be thought of as being a measure of the ``effective
dimensionality'' of the system. The functions
$f_1$ and $f_2$ for general $d$ and the crossovers of interest here
can be found in [\cite{EnvfRG}]. For $d=3$, the expressions become especially
simple, we find
$\e(z)=1-z{d\over dz}{\rm ln}
({\disp\sum_n}m^{-3})$
$$f_1(z) = 2{{\disp\sum_{n_1,n_2}}({1\over m_1^3}({1\over M}-{1\over2m_2})
+{1\over m_1M^2}({1\over m_1}+{2\over m_2}))
\over({\disp\sum_{n}}{1\over m^{3}})^{2}}\qquad \hbox{ and }\quad
f_2(z)=4{{\disp\sum_{n_1,n_2}}{1\over M^3m_1}
\over({\disp\sum_{n}}{1\over m^3})^{2}}$$
with
$m_i=(1+{4\pi^2n^2_{i}\over z^2})^{\s\frac12}$, $m_{12}=(1+{4\pi^2\over
z^2}(n_1+n_2)^2)^{\s\frac12}$, $M=m_1+m_2+m_{12}$.
We plot $\e(z)$, $f_{1}(z)$ and $f_{2}(z)$ against $\ln(1/z)$ in Figure 1.

Effective critical exponents defined as logarithmic derivatives of
the associated thermodynamic quantities with respect to  $T-T_c(L)$
at fixed $L$ for the finite size crossover and with $\G-\G_c(\beta)$
for fixed $\beta$ in the quantum problem, using the above RG
\citelabel{\Btcfss}
[\cite{DjCr},\cite{Btcfss}] can be shown to obey all the usual scaling
relations including hyperscaling. The usual dimension is
replaced by the effective dimension ${d\ef}=4-\gl$ which reflects the
changing importance of the leading irrelevant operator.
As a consequence these exponents are related to the Wilson
functions through: ${\nu\ef}={1/(2-\gft)}$, ${\eta\ef}=\gf$,
${\gamma\ef}={(2-\gf)/(2-\gft)}$, ${\alpha\ef}={(\gl-2\gft)/(2-\gft)}$,
${\beta\ef}={(2-\gl+\gf)/(4-2\gft)}$ and
${\delta\ef}={(6-\gl-\gf)/(2-\gl+\gf)}$.
Analogous effective exponents
associated with variations with respect to $L$ at fixed $T$,
and  $\beta$ at fixed $\G$ can also be
defined and computed.

We present our results in graphical form in figures 2 through 5. In all graphs
the horizontal axis is $\ln(\xi_L/L)$, the different curves correspond to $N=0$
(polymers), $N=1$ (Ising model), $N=2$ (XY-model), $N=3$ (Heisenberg model) and
$N=\i$ (spherical model). The curves represent both a four
dimensional layered geometry of thickness $L$ and a three dimensional quantum
model at $\beta=L$. The logarithmic corrections to scaling at the bulk end are
clearly visible, their magnitude is as expected from four dimensional
calculations. All curves are with the boundary condition $h=1$ at
$\ln(\xi_{L}/L)=-20$, the value of $h$ at the initial scale parameterizes
different possible crossover curves but all curves asymptote to the same form.
In Figure 2 we plot $\nu\ef$, the correlation length exponent, for $N=\i$,
${\nu\ef}\equiv{1/ ({\de}-2)}$ across the entire
crossover.
In Figure 3 we plot $\eta\ef$, the exponent which governs the fall off in
critical correlations at $T=T_c(L)$ and $\G=\G_c(\beta)$ for finite size and
quantum systems respectively.  This exponent is not a monotonic function of $N$
but attains a maximum for some value between $N=-2$ and $N=\i$, where it is
identically zero. This is the least accurate of our exponents and the peak
appears to be at $N=1$, though more accurate values for this exponent suggest
it occurs at higher values, probably $N=3$.
Figure 4 shows a plot of the effective specific heat exponent $\alpha\ef$ which
measures how the  singular part of the free energy changes as $\G$ or $T$
varies. The extra case $N=-2$ is added here, since, in the case of dimensional
crossover it is distinguishable from the Gaussian model due
to the fact that $\gl$ for the latter is zero whereas for the former it
is non-zero, being a measure of the changing effect of the leading
irrelevant operator. Across the entire crossover one  has
${\alpha\ef}=2-{\nu\ef}\de$.
Not only does one see
the change in sign of the specific heat exponent as a function
of $N$  but one also sees that the effective specific
heat exponent can change sign as a function of ${\xi_{L}/ L}$. This is quite
pronounced for the XY model which starts off positive,
increases then turns negative
at $\xi_{L}\sim 100\, L$.
It would be interesting, based on the Harris criterion
for the relevance or irrelevance of weak disorder, to see whether disorder
could change from being irrelevant to relevant as a function of size,
or temperature in the case of a quantum system.
In Figure 5 we plot $\gl=4-\de$ which also gives information about the
effective dimensionality of the system. Notice that $\gl$ is very robust to
changes in $N$, varying very little across the entire range of
$N$, $[-2,\i]$. The other effective exponents can be determined
from the effective exponent laws, which we have verified also by direct
calculation. Asymptotic values of critical exponents and associated quantities
are tabulated below.
\vfil
\centerline{\vbox{\tabskip=0pt \offinterlineskip
\def\tablerule{\noalign{\hrule}}
\halign to400pt{\strut#&\vrule#\tabskip=1em plus2em&
  #\hfil& \vrule#& #\hfil& \vrule#&
  #\hfil& \vrule#& #\hfil& \vrule#&
  #\hfil& \vrule#& #\hfil& \vrule#&
  #\hfil& \vrule#\tabskip=0pt\cr\tablerule
&&\multispan{13}\hfil Asymptotic Critical
       Exponents\hfil&\cr\tablerule
&& \omit\hidewidth N \hidewidth&&
 \omit\hidewidth $\gf$\hidewidth&&
 \omit\hidewidth $\gft$\hidewidth&&
 \omit\hidewidth $h$\hidewidth&&
 \omit\hidewidth $\gamma\ef$\hidewidth&&
 \omit\hidewidth $\nu\ef$\hidewidth&&
 \omit\hidewidth $\alpha\ef$\hidewidth&\cr\tablerule
&&\llap{-\ }2 && 0\rlap* && 0\rlap* && 1.800 && 1\rlap* && 0.5\rlap* &&
0.5\rlap* &\cr\tablerule
&&\llap{-\ }1&&0.0200&&0.145&&1.820 && 1.088&&  0.550 && 0.351 &\cr\tablerule
&&   0 && 0.0295&& 0.277 && 1.785 && 1.175&&  0.596 && 0.211 &\cr\tablerule
&&   1 && 0.0329&& 0.388 && 1.732 && 1.257&&  0.639 && 0.083 &\cr\tablerule
&&   2 && 0.0332&& 0.479 && 1.675 && 1.330&&  0.676 &&\llap{-\ }0.029
&\cr\tablerule
&&   3 && 0.0322&& 0.552 && 1.621 && 1.395&&  0.709 &&\llap{-\ }0.126
&\cr\tablerule
&&   4 && 0.0305&& 0.611 && 1.573 && 1.451&&  0.737 &&\llap{-\ }0.211
&\cr\tablerule&& $\i$&&0\rlap*&&1\rlap* &&1\rlap* && 2\rlap*&&1\rlap*&&\llap{-\
}1\rlap* &\cr\tablerule \noalign{\smallskip}
&\multispan{15}* These values are exact.\hfil\cr}}      }
All these values are in very good agreement with corresponding high
temperature series and experimental results (see
\citelabel{\LeGuZnJ}
[\cite{LeGuZnJ}] and references
therein). We  believe the entire crossover curves are of similar accuracy.

In this paper, using two loop Pad\'e
resummed perturbation theory for an environmentally friendly RG,
we presented effective critical exponents for dimensional crossover in
a four dimensional layered system with periodic boundary conditions and
quantum to classical crossover in three dimensions.
We paid special attention to polymers, Ising model, XY-model, Heisenberg
model and the spherical model. Asymptotic values for the exponents of these
systems were found to be in very good agreement with known results and
experiment. Our general formalism is applicable to a wide class of crossover
problems. Three and higher loop calculations, we believe, are quite feasible
numerically by methods similar to that of Nickel
\citelabel{\Nickel}
[\cite{Nickel}]. These should yield
effective critical exponents to the same degree of accuracy as standard
critical exponents. There is merit in pursuing such calculations as our methods
provide a direct and physical connection between exponents in different
dimensions.

\centerline{\bf REFERENCES}
\item{[\cite{Barbr}]}M. N. Barber, in Phase Transitions and Critical Phenomena,
vol.8, eds.  C. Domb and J. L. Lebowitz
(Academic Press, London 1983); CPSC vol.2,
ed. J. L. Cardy (North Holland, 1988).

\item{[\cite{Hertz}]}J. A. Hertz, Phys. Rev. {\bf B14} (1976) 1165.

\item{[\cite{ChkHaN}]}S. Chakravarty, B.I. Halperin and D.R. Nelson,
Phys. Rev. {\bf B39}
(1989) 2344.

\item{[\cite{BrLeGZnJ}]}E. Br\'ezin, J.C. Le Guillou and J. Zinn-Justin,
Phase Transitions and Critical Phenomena Vol. 6, ed.s
Domb and Green (1976).

\item{[\cite{BkNiM}]}G.A. Baker, B.G. Nickel and D.I. Meiron, Phys. Rev. {\bf
B17}
(1978) 1365.

\item{[\cite{WF}]}K.G. Wilson and M.E. Fisher,
Phys. Rev. Lett. {\bf 28} (1972) 248.

\item{[\cite{Abe}]}R. Abe, Prog. Theor. Phys. {\bf 48} (1972) 1414.

\item{[\cite{Parisi}]}G. Parisi, J. Stat. Phys. {\bf 23} (1980) 49.

\item{[\cite{AG}]}D.J. Amit and Y.Y. Goldschmidt, Ann. Phys., {\bf 114} (1978)
356.

\item{[\cite{NeDo}]} D.R. Nelson  and E. Domany, Phys. Rev. {\bf B13} (1976)
236;
P. Seglar and M.E. Fisher, J. Phys. {\bf C13} (1980) 6613.

\item{[\cite{DjCr}]} Denjoe O'Connor and C.R. Stephens, Proc. Roy. Soc. {\bf A}
(1993) to be published.

\item{[\cite{FSch}]}E. Frey and F. Schwabl, Phys. Rev. B, {\bf 42} (1990) 8261.

\item{[\cite{Chr}]} C.R. Stephens, J. of Magnetism and Magnetic Materials,
104-107 (1992) 297.

\item{[\cite{Schm}]}D. Schmeltzer, Phys. Rev. {\bf B32}, 7512 (1985).

\item{[\cite{Lawrie}]}I. D. Lawrie, J. Phys. {\bf C 11}, (1978), 3857.

\item{[\cite{BrZnJ}]}E. Br\'ezin and J. Zinn-Justin,
Nucl. Phys. {\bf B257}\ [FS14](1985) 867;
J. Rudnick, H. Guo and D. Jasnow, Jour. Stat. Phys. {\bf 41} (1985) 353.

\item{[\cite{EnvfRG}]}Denjoe O'Connor and C.R. Stephens, ``Environmentally
Friendly
Renormalization'', Utrecht/DIAS preprint THU-93/14.

\item{[\cite{Sigmod}]}Denjoe O'Connor and C.R. Stephens,
``Dimensional Crossover in the
Non-linear $\sigma$ model'',  Utrecht/DIAS preprint.

\item{[\cite{NPhyOrig}]}Denjoe O'Connor and C. R. Stephens,
Nucl. Phys. {\bf B360} (1991) 297.

\item{[\cite{Btcfss}]}F. Freire, D. O'Connor and C.R. Stephens,
Dimensional Crossover and Finite Size Scaling Below $T_c$,
Univ. Utrecht preprint THU 92/36.

\item{[\cite{LeGuZnJ}]}J.C. Le Guillou and J. Zinn-Justin,
Phys. Rev. {\bf B21} (1980) 3976.

\item{[\cite{Nickel}]}B.G. Nickel, J. Math. Phys. {\bf 19} (1978) 542.

\end